\documentstyle[aps,epsfig,12pt]{revtex}

\tightenlines

\begin{document}
ADP-04-19/T601

JLAB-THY-04-26
\begin{center}
\vspace*{2cm} {\Large {\bf Liquid-gas phase transition and Coulomb instability
of asymmetric nuclear systems}}\\[0pt]
\vspace*{1cm} P. Wang$^a$, D. B. Leinweber$^a$, A. W. Thomas$^{a,b}$ and A. G. Williams$^a$ \\[0pt]
\vspace*{0.2cm} {\it $^a$Special Research Center for the Subatomic Structure of
Matter (CSSM) and Department of Physics, University of Adelaide 5005, Australia } \\[0pt]
\vspace*{0.2cm} {\it $^b$Jefferson Laboratory, 12000 Jefferson Ave., Newport News, VA 23606 USA}\\[0pt]
\end{center}

\begin{abstract}
We use a chiral $SU(3)$ quark mean field model to study the properties
of nuclear
systems at finite temperature. The liquid-gas phase transition of
symmetric and asymmetric nuclear matter is discussed. For two formulations
of the model the
critical temperature, $T_c$, for symmetric nuclear matter is found
to be 15.8 MeV
and 17.9 MeV. These values are consistent with those
derived from recent experiments. The limiting temperatures for finite
nuclei are in good agreement with the experimental points.
\end{abstract}

\bigskip

\leftline{PACS number(s): 21.65.+f; 12.39.-x; 11.30.Rd}
\bigskip
\leftline{{\bf Keywords: Liquid-gas Phase Transition, Coulomb Instability,
Chiral Symmetry,}}

{\bf ~~~~~~~~~~ Nuclear Matter, Quark Mean Field}

\section{Introduction}

The determination of the properties of hadronic matter at finite
temperature and density is a fundamental problem in nuclear
physics. In particular, the study of liquid-gas phase transition
in medium energy heavy-ion collisions is of considerable interest.
Many intermediate-energy collision experiments have been performed
\cite{Suraud} to investigate the
unknown features of the highly excited or hot nuclei formed in
collisions \cite{Panagiotou,Chen}. Theoretically, much effort has
been devoted to studying the equation of state for nuclear matter
and to discussing the critical temperature, $T_c$.
The calculated critical
temperature of symmetric nuclear matter lies in the range 13 - 24 MeV
for various phenomenological models \cite{Bonche}-\cite{Kolomietz}.
Glendenning \cite{Glendenning} first discussed the phase transition
with more than one conserved charge
and applied the method to the possible transition to quark
matter in the core of a neutron star.
M\"uller and Serot \cite{Muller} discussed
asymmetric nuclear matter using the stability conditions on the
free energy, conservation laws and the Gibbs criterion for the
liquid-gas phase transition. The liquid-gas phase transition
of asymmetric nuclear matter has also been discussed in
effective chiral models \cite{Wang1,Qian}.
It was found that the critical temperature decreases with increasing
asymmetry parameter, $\alpha$.

For finite nuclei, there is another
temperature which is called
the limiting temperature, $T_{lim}$, as pointed out by
Levit and Bonche \cite{Levit}.
Below the limiting temperature, nuclei can exist in equilibrium with
the surrounding vapor. When the temperature is higher than $T_{lim}$
the nuclei are unstable and will fragment. This is called Coulomb
instability. The size effect and Coulomb interaction are important
in the determination of the limiting temperature, resulting in a lower
limiting temperature compared with that for infinite nuclear matter
\cite{Pawlowski}. Recently, Natowitz $et$ $al.$ obtained the
limiting temperature by using a number of different experimental
measurements \cite{Natowitz1}. From these observations the authors
extracted the critical temperature of infinite nuclear matter
$T_c=16.6\pm 0.86$ MeV \cite{Natowitz2}.
Their results show that the limiting temperature is in good agreement with
the previous calculations employing either a chiral symmetric model
\cite{Zhang2} or the Gogny interaction \cite{Zhang1}.

To study the properties of hadronic matter, we need
phenomenological models since QCD cannot yet be used directly. The
symmetries of QCD can be used to determine largely how the hadrons
interact with each other. On this basis, models based on
$SU(2)_{L}\times SU(2)_{R}$ symmetry and scale invariance were
proposed. These effective models have been widely used in recent
years to investigate nuclear matter and finite nuclei, both at
zero temperature and at finite temperature
\cite{Zhang2}-\cite{Furnstahl}. Papazoglou $et$ $al.$ extended the
chiral effective models to $SU(3)_{L}\times SU(3)_{R}$ including
the baryon octets\cite{Papazoglou1,Papazoglou2}. As well as the
models based on the hadron degrees of freedom, there are
additional models based on quark degrees of freedom, such as the
quark meson coupling model \cite{Guichon,Kazuo}, the cloudy bag
model \cite{Thomas}, the NJL model \cite{Bentz} and the quark mean
field model \cite{Toki}. Recently, we proposed a chiral $SU(3)$
quark mean field model and applied to investigate hadronic matter
and quark matter \cite{Wang3}-\cite{Wang6}. This model is very
successful in describing the properties of nuclear matter
\cite{Wang3}, strange matter \cite{Wang4,Wang5}, finite nuclei and
hypernuclei \cite{Wang6} at zero temperature. A successful model
should describe well the properties of nuclear matter, not only at
zero temperature but also at finite temperature. In this paper, we
will apply the chiral $SU(3)$ quark mean field model to study the
liquid-gas phase transition and Comloub instability of asymmetric
nuclear system and compare our results with the recent
experimental analysis.

The paper is organized as follows. The model is introduced in section II. In
section III, we apply it to investigate nuclear matter
at finite temperature. The numerical results are discussed in section IV and
section V summarises the main results.

\section{The model}

Our considerations are based on the chiral $SU(3)$ quark mean field model
(for details see Refs.~\cite{Wang4,Wang6}), which contains quarks and mesons
as basic degrees of freedom. Quarks are confined into baryons by an
effective potential. The quark meson interaction and meson self-interaction
are based on $SU(3)$ chiral symmetry. Through the mechanism of spontaneous
chiral symmetry breaking, the resulting constituent quarks and mesons (except for
the pseudoscalars) obtain masses. The introduction of an explicit symmetry
breaking term in the meson self-interaction generates the masses of
the pseudoscalar mesons which satisfy the partially conserved axial-vector current
(PCAC) relation. The explicit symmetry breaking term of the
quark meson interaction leads in turn to reasonable hyperon potentials in
hadronic matter. For completeness, we introduce the main concepts of the
model in this section.

In the chiral limit, the quark field $q$ can be
split into left and right-handed parts $q_{L}$ and $q_{R}$:
$q\,=\,q_{L}\,+\,q_{R}$. Under $SU(3)_{L}\times SU(3)_{R}$ they
transform as
\begin{equation}
q_{L}^{\prime }\,=\,L\,q_{L},~~~~~q_{R}^{\prime }\,=\,R\,q_{R}\,.
\end{equation}
The spin-0 mesons are written in the compact form
\begin{equation}
M(M^{+})=\Sigma \pm i\Pi =\frac{1}{\sqrt{2}}\sum_{a=0}^{8}\left( \sigma
^{a}\pm i\pi ^{a}\right) \lambda ^{a},
\end{equation}
where $\sigma ^{a}$ and $\pi ^{a}$ are the nonets of scalar and
pseudoscalar mesons, respectively, $\lambda ^{a}(a=1,...,8)$ are the
Gell-Mann matrices, and $\lambda ^{0}=\sqrt{\frac{2}{3}}\,I$.
The alternatives, plus and minus signs correspond to $M$ and $M^{+}$.
Under chiral $SU(3)$ transformations, $M$ and $M^{+}$ transform as
$M\rightarrow M^{\prime }=LMR^{+}$ and $M^{+}\rightarrow
M^{+^{\prime }}=RM^{+}L^{+}$. In a similar way, the spin-1 mesons
are introduced through:
\begin{equation}
l_{\mu }(r_{\mu })=\frac{1}{2}\left( V_{\mu }\pm A_{\mu }\right)
= \frac{1}{2\sqrt{2}}\sum_{a=0}^{8}\left( v_{\mu }^{a}\pm a_{\mu }^{a}
\right) \lambda^{a}
\end{equation}
with the transformation properties:
$l_{\mu }\rightarrow l_{\mu }^{\prime }=Ll_{\mu }L^{+}$,
$r_{\mu }\rightarrow r_{\mu }^{\prime }=Rr_{\mu }R^{+}$.
The matrices $\Sigma$, $\Pi$,
$V_{\mu }$ and $A_{\mu }$ can be written in
a form where the physical states are explicit. For the scalar and vector
nonets, we have the expressions
\begin{eqnarray}
\Sigma&=&\frac1{\sqrt{2}}\sum_{a=0}^8\sigma^a\lambda^a=\left(
\begin{array}{lcr}
\frac1{\sqrt{2}}\left(\sigma+a_0^0\right) & a_0^+ & K^{*+} \\
a_0^- & \frac1{\sqrt{2}}\left(\sigma-a_0^0\right) & K^{*0} \\
K^{*-} & \bar{K}^{*0} & \zeta
\end{array}
\right),
\end{eqnarray}
\begin{eqnarray}
V_\mu&=&\frac1{\sqrt{2}}\sum_{a=0}^8 v_\mu^a\lambda^a=\left(
\begin{array}{lcr}
\frac1{\sqrt{2}}\left(\omega_\mu+\rho_\mu^0\right) & \rho_\mu^+ &
K_\mu^{*+}\\
\rho_\mu^- & \frac1{\sqrt{2}}\left(\omega_\mu-\rho_\mu^0\right) &
K_\mu^{*0}\\
K_\mu^{*-} & \bar{K}_\mu^{*0} & \phi_\mu
\end{array}
\right).
\end{eqnarray}
Pseudoscalar and pseudovector nonet mesons can be written in a similar
fashion.

The total effective Lagrangian has the form:
\begin{eqnarray}
{\cal L}_{{\rm eff}} \, = \, {\cal L}_{q0} \, + \, {\cal L}_{qM} \, + \,
{\cal L}_{\Sigma\Sigma} \, + \, {\cal L}_{VV} \, + \, {\cal L}_{\chi SB}\,
+ \, {\cal L}_{\Delta m_s} \, + \, {\cal L}_{h}, + \, {\cal L}_{c},
\end{eqnarray}
where ${\cal L}_{q0} =\bar q \, i\gamma^\mu \partial_\mu \, q$ is the
free part for massless quarks. The quark-meson interaction ${\cal L}_{qM}$
can be written in a chiral $SU(3)$ invariant way as
\begin{eqnarray}
{\cal L}_{qM}=g_s\left(\bar{\Psi}_LM\Psi_R+\bar{\Psi}_RM^+\Psi_L\right)
-g_v\left(\bar{\Psi}_L\gamma^\mu l_\mu\Psi_L+\bar{\Psi}_R\gamma^\mu
r_\mu\Psi_R\right)~~~~~~~~~~~~~~~~~~~~~~~  \nonumber \\
=\frac{g_s}{\sqrt{2}}\bar{\Psi}\left(\sum_{a=0}^8\sigma_a\lambda_a
+i\sum_{a=0}^8\pi_a\lambda_a\gamma^5\right)\Psi -\frac{g_v}{2\sqrt{2}}
\bar{\Psi}\left(\sum_{a=0}^8\gamma^\mu v_\mu^a\lambda_a
-\sum_{a=0}^8\gamma^\mu\gamma^5 a_\mu^a\lambda_a\right)\Psi.
\end{eqnarray}
In the mean field approximation, the chiral-invariant scalar meson
${\cal L}_{\Sigma\Sigma}$ and vector meson ${\cal L}_{VV}$
self-interaction terms are written as~\cite{Wang4,Wang6}
\begin{eqnarray}
{\cal L}_{\Sigma\Sigma} &=& -\frac{1}{2} \, k_0\chi^2
\left(\sigma^2+\zeta^2\right)+k_1 \left(\sigma^2+\zeta^2\right)^2
+k_2\left(\frac{\sigma^4}2 +\zeta^4\right)+k_3\chi\sigma^2\zeta
\nonumber \\ \label{scalar}
&&-k_4\chi^4-\frac14\chi^4 {\rm ln}\frac{\chi^4}{\chi_0^4} +\frac{\delta}
3\chi^4 {\rm ln}\frac{\sigma^2\zeta}{\sigma_0^2\zeta_0}, \\
{\cal L}_{VV}&=&\frac{1}{2} \, \frac{\chi^2}{\chi_0^2} \left(
m_\omega^2\omega^2+m_\rho^2\rho^2+m_\phi^2\phi^2\right)+g_4
\left(\omega^4+6\omega^2\rho^2+\rho^4+2\phi^4\right), \label{vector}
\end{eqnarray}
where $\delta = 6/33$; $\sigma_0$, $\zeta_0$ and $\chi_0$ are the vacuum
expectation values of the corresponding mean fields $\sigma$, $\zeta$
and $\chi$. The Lagrangian ${\cal L}_{\chi SB}$ generates nonvanishing
masses for the pseudoscalar mesons
\begin{equation}\label{ecsb}
{\cal L}_{\chi SB}=\frac{\chi^2}{\chi_0^2}\left[m_\pi^2F_\pi\sigma +
\left(
\sqrt{2} \, m_K^2F_K-\frac{m_\pi^2}{\sqrt{2}} F_\pi\right)\zeta\right],
\end{equation}
leading to a nonvanishing divergence of the axial currents which in turn satisfy
the relevant PCAC relations for $\pi$ and $K$ mesons. Pseudoscalar and scalar
mesons as well as the dilaton field, $\chi$, obtain mass terms by
spontaneous breaking of chiral symmetry in the Lagrangian (\ref{scalar}).
The masses of the $u$, $d$ and $s$ quarks are generated by the vacuum
expectation values of the two scalar mesons $\sigma$ and $\zeta$.
To obtain the correct constituent mass of the strange quark, an
additional mass term has to be added:
\begin{eqnarray}
{\cal L}_{\Delta m_s} = - \Delta m_s \bar q S q
\end{eqnarray}
where $S \, = \, \frac{1}{3} \, \left(I - \lambda_8\sqrt{3}\right) =
{\rm diag}(0,0,1)$ is the strangeness quark matrix.
Based on these mechanisms,
the quark constituent masses are finally given by
\begin{eqnarray}
m_u=m_d=-\frac{g_s}{\sqrt{2}}\sigma_0
\hspace*{.5cm} \mbox{and} \hspace*{.5cm}
m_s=-g_s \zeta_0 + \Delta m_s,
\end{eqnarray}
where $g_s$ and $\Delta m_s$ are chosen to yield the constituent
quark mass in vacuum -- we use $m_u=m_d=313$ MeV and $m_s=490$ MeV. In
order to obtain reasonable hyperon potentials in hadronic matter,
it has been found necessary to include an additional coupling between
strange quarks and the scalar mesons
$\sigma$ and $\zeta$ \cite{Wang4}:
\begin{eqnarray}
{\cal L}_h \, = \, (h_1 \, \sigma \, + \, h_2 \, \zeta) \, \bar{s} s \, .
\end{eqnarray}
In the quark mean field model, quarks are confined in baryons
by the Lagrangian ${\cal L}_c=-\bar{\Psi} \, \chi_c \, \Psi$ (with $\chi_c$
given in Eq. (\ref{Dirac}), below).
The Dirac equation for a quark field $\Psi_{ij}$ under the additional
influence of the meson mean fields is given by
\begin{equation}
\left[-i\vec{\alpha}\cdot\vec{\nabla}+\chi_c(r)+\beta m_i^*\right]
\Psi_{ij}=e_i^*\Psi_{ij}, \label{Dirac}
\end{equation}
where $\vec{\alpha} = \gamma^0 \vec{\gamma}$\,, $\beta = \gamma^0$\,,
the subscripts $i$ and $j$ denote the quark $i$ ($i=u, d, s$)
in a baryon of type $j$ ($j=N, \Lambda, \Sigma, \Xi$)\,;
$\chi_c(r)$ is a confinement potential, i.e. a static potential
providing confinement of quarks by meson mean-field configurations.
The quark effective mass, $m_i^*$, and energy, $e_i^*$, are defined as
\begin{equation}
m_i^*=-g_\sigma^i\sigma - g_\zeta^i\zeta+m_{i0}
\end{equation}
and
\begin{equation}
e_i^*=e_i-g_\omega^i\omega-g_\phi^i\phi-g_\rho^i\rho\,,
\end{equation}
where $e_i$ is the energy of the quark under the influence of
the meson mean fields. Here $m_{i0} = 0$ for $i=u,d$ (nonstrange quark)
and $m_{i0} = \Delta m_s = 29$~MeV for $i=s$ (strange quark).
Using the solution of the Dirac
equation~(\ref{Dirac}) for the quark energy $e_i^*$
it has been common to define
the effective mass of the baryon $j$ through the ans\"atz:
\begin{eqnarray}
M_j^*=\sqrt{E_j^{*2}- <p_{j \, cm}^{*2}>} \label{square}\,,
\end{eqnarray}
where $E_j^*=\sum_in_{ij}e_i^*+E_{j \, spin}$ is the baryon energy and
$<p_{j \, cm}^{*2}>$ is the subtraction of the contribution
to the total energy associated with spurious center of mass
motion. In the expression for the baryon energy $n_{ij}$ is the number
of quarks with flavor $"i"$ in a baryon
with flavor $j$, with $j = N \, \{p, n\}\,,
\Sigma \, \{\Sigma^\pm, \Sigma^0\}\,, \Xi \,\{\Xi^0, \Xi^-\}\,,
\Lambda\,$  and $E_{j \, spin}$ is the correction
to the baryon energy which is determined from a fit to the data for
baryon masses.

There is an alternative way to remove the spurious c.\ m.\ motion and
determine the effective baryon masses. In Ref.~\cite{Guichon2},
the removal of the spurious c.\ m.\ motion for three quarks moving in
a confining, relativistic oscillator potential was studied in some
detail. It was found that when an external scalar potential was
applied, the effective mass obtained from the interaction Lagrangian
could be written as
\begin{eqnarray}
M_j^*=\sum_in_{ij}e_i^*-E_j^0 \label{linear}\,,
\end{eqnarray}
where $E_j^0$ was calculated to be only very weakly dependent on the
external field strength.
We therefore use Eq.~(\ref{linear}), with $E_j^0$ a
constant, independent of the density, which is adjusted to give a
best fit to the free baryon masses.

Using the square root ans\"atz for the effective baryon
mass, Eq.~(\ref{square}), the confining potential
$\chi_{c}$ is chosen as a combination of scalar
(S) and scalar-vector (SV) potentials as in Ref.~\cite{Wang4}:
\begin{eqnarray}
\chi_{c}(r)=\frac12 [\,\chi_{c}^{\rm S}(r)
                         + \chi_{c}^{\rm SV}(r)\,]
\end{eqnarray}
with
\begin{eqnarray}
\chi_{c}^{\rm S}(r)=\frac14 k_{c} \, r^2 \,,
\end{eqnarray}
and
\begin{eqnarray}
\chi_{c}^{\rm SV}(r)=\frac14 k_{c} \, r^2(1+\gamma^0) \,.
\end{eqnarray}
On the other hand, using the linear definition of effective baryon
mass, Eq.~(\ref{linear}), the confining potential
$\chi_{c}$ is chosen to be the purely scalar potential $\chi_{c}^{\rm S}(r)$.
The coupling $k_{c}$ is taken as
$k_{c} = 1$ (GeV fm$^{-2})$, which yields root-mean-square baryon
charge radii (in the absence
of a pion cloud \cite{Hackett}) around 0.6 fm.

\section{nuclear matter at finite temperature}

Based on the previously defined quark mean field model
the Lagrangian density for nuclear matter is written as
\begin{eqnarray}
{\cal L}&=&\bar{\psi}(i\gamma^\mu\partial_\mu-M_N^*)\psi
+\frac12\partial_\mu\sigma\partial^\mu\sigma+\frac12
\partial_\mu\zeta\partial^\mu\zeta+\frac12\partial_\mu
\chi\partial^\mu\chi-\frac14F_{\mu\nu}F^{\mu\nu} -
\frac14 \rho_{\mu\nu}\rho^{\mu\nu}\nonumber \\
&&-g_\omega\bar{\psi}\gamma_\mu\psi\omega^\mu -g_\rho\bar{\psi}
_B\gamma_\mu\tau_3\psi\rho^\mu +{\cal L}_M, \label{bmeson}
\end{eqnarray}
where
\begin{equation}
F_{\mu\nu}=\partial_\mu\omega_\nu-\partial_\nu\omega_\mu
\hspace*{.5cm} \mbox{and} \hspace*{.5cm}
\rho_{\mu\nu}=\partial_\mu\rho_\nu-\partial_\nu\rho_\mu.
\end{equation}
The term ${\cal L}_M$ represents the interaction between mesons which
includes the scalar meson self-interaction ${\cal L}_{\Sigma\Sigma}$,
the vector meson self-interaction
${\cal L}_{VV}$ and the explicit chiral symmetry breaking term
${\cal L}_{\chi SB}$, all defined previously.
The Lagrangian includes the scalar mesons
$\sigma$, $\zeta$ and $\chi$, and the vector mesons $\omega$ and $\rho$.
The interactions between quarks and scalar mesons result in the effective
nucleon mass $M_N^*$, The interactions between quarks and
vector mesons generate the nucleon-vector meson interaction terms of
equation (\ref{bmeson}).
The corresponding vector coupling constants $g_\omega$ and $g_\rho$ are
baryon dependent and satisfy the $SU(3)$ relationship:
$g_\rho^p=-g_\rho^n=\frac13 g_\omega^p=\frac13 g_\omega^n$.

At finite temperature and density, the thermodynamic potential is
defined as
\begin{eqnarray}
\Omega = - \frac{k_{B}T}{(2\pi)^3}\sum_{N=p,n}
\int_0^\infty d^3\overrightarrow{k}\biggl\{{\rm ln}
\left( 1+e^{-(E_N^{\ast}(k) - \nu _N)/k_{B}T}\right)
+ {\rm ln}\left( 1+e^{-(E_N^{\ast }(k)+\nu_N)/k_{B}T}
\right) \biggr\} -{\cal L}_{M},
\end{eqnarray}
where $E_N^{\ast }(k)=\sqrt{M_N^{\ast 2}+\overrightarrow{k}^2}$.
The quantity $\nu _N$ is related to the usual chemical potential,
$\mu _N$, by $\nu _N =\mu _N-g_{\omega }^N\omega -g_{\rho }^N\rho$.
The energy per unit volume and the pressure of the system are respectively
$\varepsilon =\Omega -\frac1T
\frac{\partial\Omega}{\partial T}+\nu _N\rho_N$ and $p=-\Omega $,
where $\rho_N$ is the baryon density.

The mean field equation for meson $\phi _{i}$ is obtained by the
formula $\partial \Omega/\partial \phi_i=0$. For example,
the equations for $\sigma$, $\zeta$ are deduced as:
\begin{eqnarray}\label{eq_sigma}
k_{0}\chi ^{2}\sigma
-4k_{1}\left( \sigma ^{2}+\zeta ^{2}\right) \sigma -2k_{2}\sigma
^{3}-2k_{3}\chi \sigma \zeta -\frac{2\delta }{3\sigma }\chi ^{4}
+\frac{\chi^{2}}{\chi _{0}^{2}}m_{\pi }^{2}F_{\pi }  \nonumber \\
-\left( \frac{\chi }{\chi _{0}}\right) ^{2}m_{\omega }\omega ^{2}\frac{
\partial m_{\omega }}{\partial \sigma }
-\left( \frac{\chi }{\chi _{0}}\right) ^{2}m_{\rho }\rho ^{2}\frac{
\partial m_{\rho }}{\partial \sigma }
+\frac{\partial M_N^{\ast }}{\partial \sigma } <\bar{\psi}\psi>=0,
\end{eqnarray}
\begin{eqnarray}\label{eq_zeta}
k_{0}\chi ^{2}\zeta -4k_{1}\left(\sigma ^{2}+\zeta ^{2}\right)
\zeta -4k_{2}\zeta ^{3}-k_{3}\chi \sigma ^{2} -
\frac{\delta }{3\zeta }\chi ^{4}+\frac{\chi ^{2}}{\chi _{0}^{2}}
\left( \sqrt{2}m_{k}^{2}F_{k}-\frac{1}{\sqrt{2}}m_{\pi }^{2}F_{\pi }
\right)=0
\end{eqnarray}
where
\begin{equation}
<\bar{\psi}\psi>=\frac{1}{\pi ^{2}}\int_{0}^{\infty}
dk \frac{k^{2}M_{N}^{\ast }}{E^*(k)}
\left[n_n(k)+\bar{n}_n(k)+n_p(k)+\bar{n}_p(k)\right].
\end{equation}
In the above equation, $n_q(k)$ and $\bar{n}_q(k)$ are the nucleon
and antinucleon distributions, respectively, expressed as
\begin{equation}
n_q(k)=\frac{1}{exp\left[\left(E^*(k)-\nu_q\right)/k_B T\right]+1}
\end{equation}
and
\begin{equation}
n_q(k)=\frac{1}{exp\left[\left(E^*(k)+\nu_q\right)/k_B T\right]+1}
~~~~(q=n,p).
\end{equation}
The equations for the vector mesons, $\omega$ and $\rho$, are:
\begin{equation}
\frac{\chi^2}{\chi_0^2}m_\omega^2\omega+4g_4\omega^3+12g_4\omega\rho^2
=g_\omega^N(\rho_p+\rho_n),
\end{equation}
\begin{equation}
\frac{\chi^2}{\chi_0^2}m_\rho^2\rho+4g_4\rho^3+12g_4\omega^2\rho
=\frac13 g_\omega^N(\rho_p-\rho_n),
\end{equation}
where $\rho_p$ and $\rho_n$ are the proton and neutron densities,
expressed as
\begin{equation}
\rho_q=\frac{1}{\pi ^{2}}\int_{0}^{\infty}
dk k^2\left[n_q(k)-\bar{n}_q(k)\right] ~~~~ (q=p,n).
\end{equation}
In order to describe asymmetric nuclear matter, one can introduce
the asymmetry parameter $\alpha$ which is defined as
\begin{equation}
\alpha=\frac{\rho_n-\rho_p}{\rho_N},
\end{equation}
where $\rho_N=\rho_n+\rho_p$.
For symmetric matter $\alpha=0$, while for neutron matter
$\alpha=1$.

Let us now discuss the liquid-gas phase transition. For
asymmetric nuclear matter we follow the thermodynamic approach
of Refs. \cite{Glendenning} and \cite{Muller}. The system will be
stable against separation into two phases if the free energy of
a single phase is lower than the free energy in all two-phase
configurations. This requirement can be formulated as
\cite{Muller}
\begin{equation}
F(T,\rho)<(1-\lambda)F(T,\rho^\prime)+\lambda F(T,\rho^{\prime\prime}),
\end{equation}
with
\begin{equation}
\rho=(1-\lambda)\rho^\prime+\lambda\rho^{\prime\prime}, ~~~~~ 0<\lambda<1,
\end{equation}
where $F$ is the Helmholtz free energy per unit volume. The two phases
are denoted by a prime and a double prime. The stability
condition implies the following set of inequalities:
\begin{equation}
\rho\left(\frac{\partial p}{\partial \rho}\right)_{T,\alpha}>0,
\end{equation}
\begin{equation}
\left(\frac{\partial \mu_p}{\partial \alpha}\right)_{T,p}<0 ~~~~
\text{or}
~~~~ \left(\frac{\partial \mu_n}{\partial \alpha}\right)_{T,p}>0.
\end{equation}
If one of the stability conditions is violated, a system with two
phases is energetically favored. The phase coexistence is
governed by the Gibbs conditions:
\begin{equation}\label{eq_Gibbsmu}
\mu_q^\prime(T,\rho^\prime)=\mu_q^{\prime\prime}(T,\rho^{\prime\prime}), ~~~~ (q=n,p),
\end{equation}
\begin{equation}\label{eq_Gibbsp}
p^\prime(T,\rho^\prime)=p^{\prime\prime}(T,\rho^{\prime\prime}),
\end{equation}
where the temperature is the same in the two phases.

\section{Numerical results and discussions}
\subsection{Liquid-gas phase transition}

The parameters in this model are determined by the meson masses
in vacuum and the properties of nuclear matter which were listed in table I
of Ref.~\cite{Wangcssm}. We first discuss the liquid-gas
phase transition of symmetric nuclear matter. In Fig.~1, we show
the pressure of the system versus nucleon density at different
temperatures using the square root ans\"atz for the effective
nucleon mass (Eq. (\ref{square})). At low temperature, the
pressure first increases and then decreases with increasing
density. The $p-\rho_N$ isotherms exhibit the form of two phase
coexistence, with an unphysical region for each. At temperature
$T=15.82$ MeV, there appears a point of inflection, where
$\partial p/\partial\rho_N=0$, $\partial ^2p/\partial \rho_N^2=0$.
This temperature is called the critical temperature. Symmetric
nuclear matter can only be in gas phase above this temperature.
The pressure versus nucleon density with the linear definition of
effective nucleon mass (Eq. (\ref{linear})) is shown in Fig.~2. In
this case, the critical temperature is 17.9 MeV. In both cases,
the calculated critical temperature is close to the recent result
$T_c=16.6\pm 0.86$ MeV, which was obtained by Natowitz $et$ $al.$
\cite{Natowitz2}. The critical temperature calculated with the
Walecka model is larger than
20 MeV. This large $T_c$ is partially caused by the large incompressibility
modulus K ($\simeq$ 540 MeV). For the same reason the stiff Skyrme interaction
SK1 gives a large $T_c$, which is close to 20 MeV, while the soft Skyrme
interaction SKM$^*$ gives a small $T_c$ \cite{Song1}. The QMC model and the
effective model based on $SU(2)$ chiral symmetry provide reasonable values
of $T_c$, around 15-16 MeV \cite{Wang1,Songhq}.

\begin{center}
\begin{figure}
\centering{\
\epsfig{figure=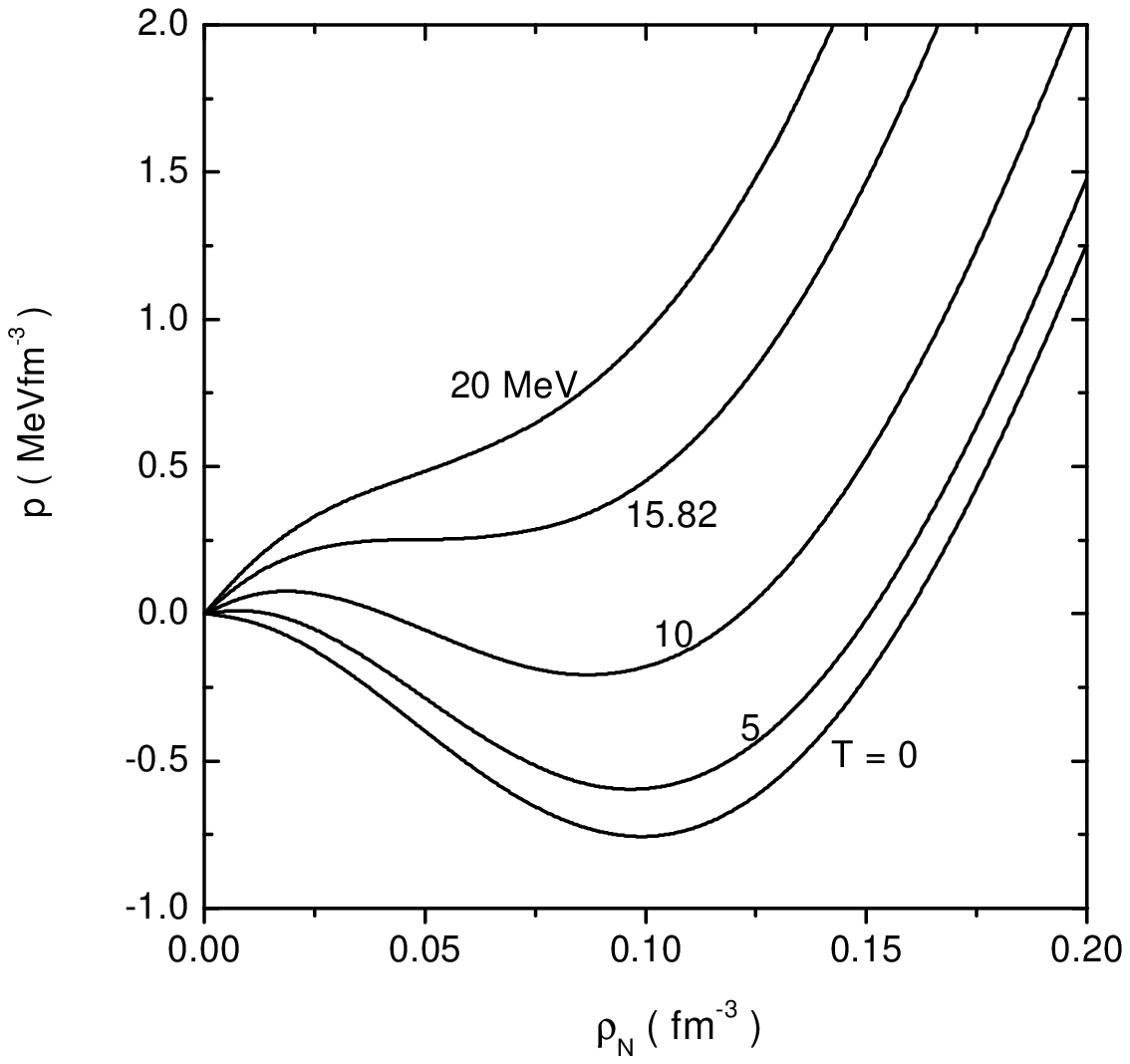,height=8cm}
}
\caption{The pressure of symmetric nuclear matter $p$ versus nucleon
density, $\rho_N$, at different temperatures with the square root
ans\"atz of effective nucleon mass.}
\end{figure}
\end{center}

For the asymmetric case,
the situation is more complicated. One cannot get the critical
temperature from the $p-\rho_N$ isotherms. The chemical potentials of
the proton and neutron are different. We show the chemical
potential versus asymmetric parameter $\alpha$ at temperature
$T=10$ MeV and pressure $p=0.12$ MeVfm$^{-3}$ with the square
root definition of effective nucleon mass in Fig.~3 (For convenience,
we use the reduced chemical potential which is defined as
$\widetilde{\mu}_N=\mu_N-M_N$.) The solid and dashed lines are for
proton and neutron respectively.
The Gibbs equations (\ref{eq_Gibbsmu}) and (\ref{eq_Gibbsp})
for phase equilibrium demand equal
pressure and chemical potentials for two phases with different
concentrations. The desired solution can be found by means of
the geometrical construction shown in Fig.~3, which guarantees the
same pressure and chemical potentials for protons and neutrons
in the two phases with different asymmetry parameter $\alpha$.

\begin{center}
\begin{figure}
\centering{\
\epsfig{figure=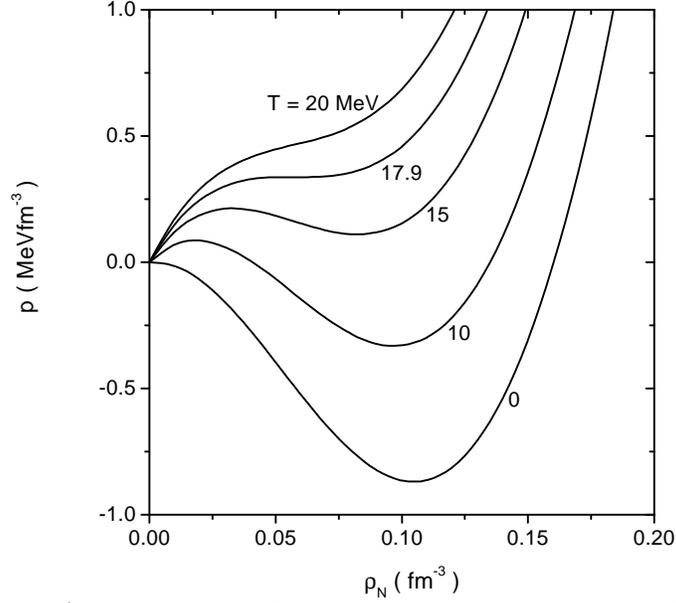,height=8cm}
}
\caption{The pressure of symmetric nuclear matter $p$ versus nucleon
density $\rho_N$ at different temperatures with the linear
definition of effective nucleon mass.}
\end{figure}
\end{center}

The pairs of solutions found using the method just described, yield
a binodal curve which is
shown in Fig.~4. There is a critical point where the pressure
is about 0.205 MeV-fm$^{-3}$ and the corresponding asymmetry
parameter is around 0.7. The two phases have the same $\alpha$ and
therefore the same density at this point. The binodal curve is
divided into two branches by the critical point. One branch
corresponds to the high density (liquid) phase, the other
corresponds to the low density (gas) phase. Assume the system is
initially prepared in the low density (gas) phase with
$\alpha=0.55$. When the pressure increases to some value, the
two-phase region is encountered at point A and a liquid phase at B
with a low $\alpha$ begins to emerge. As the system is compressed,
the gas phase evolves from point A to C, while the liquid phase
evolves from B to D. If the pressure of the system continues to
increase, the system will leave the two-phase region at point
D. The gas phase disappears and the system is entirely in the
liquid phase. This kind of phase transition is different from the
normal first order phase transition where the pressure remains
constant during phase transition. If the initial asymmetric
nucleon gas is larger than some value, the system enters and
leaves the two phase region on the same branch. For example, the
system becomes unstable at point A$^\prime$ and a liquid  phase
with a higher nucleon density begins to emerge at B$^\prime$. The
system is compressed at a fixed total $\alpha$, with the gas phase
evolving from A$^\prime$ to C$^\prime$ and the liquid phase from
B$^\prime$ to D$^\prime$.The system leaves the two-phase region
point C$^\prime$ which is still in the original gas phase.
Therefore, for a given temperature, if the total asymmetry
parameter of the system is larger than a critical value, the
system cannot change completely into the liquid phase, however large
the pressure. In other words,
for a system with a fixed asymmetric parameter $\alpha$,
there exists a critical temperature, above which the system can
only be in the gas phase at any pressure.

\begin{center}
\begin{figure}[hbt]
\centering{\
\epsfig{figure=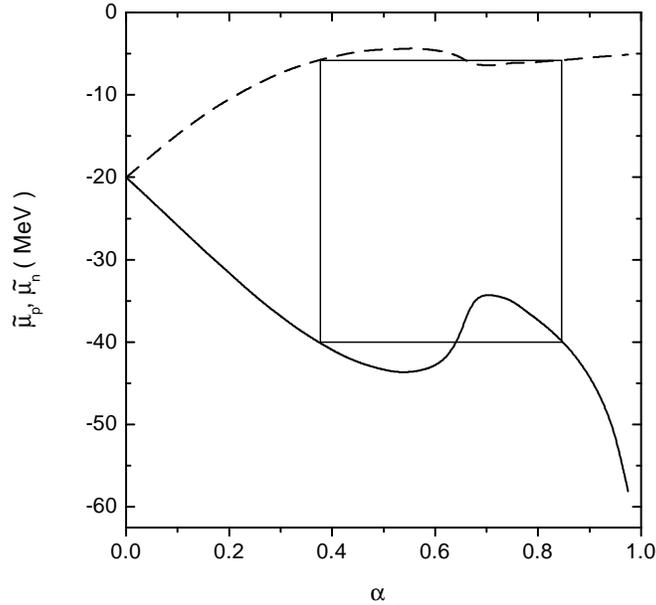,height=8cm}
}
\caption{Geometrical construction used to obtain the chemical
potentials and asymmetric parameters in the two-phase coexistence
at temperature $T=10$ MeV and $p=0.12$ MeVfm$^{-3}$.}
\end{figure}
\end{center}

\begin{center}
\begin{figure}[hbt]
\centering{\
\epsfig{figure=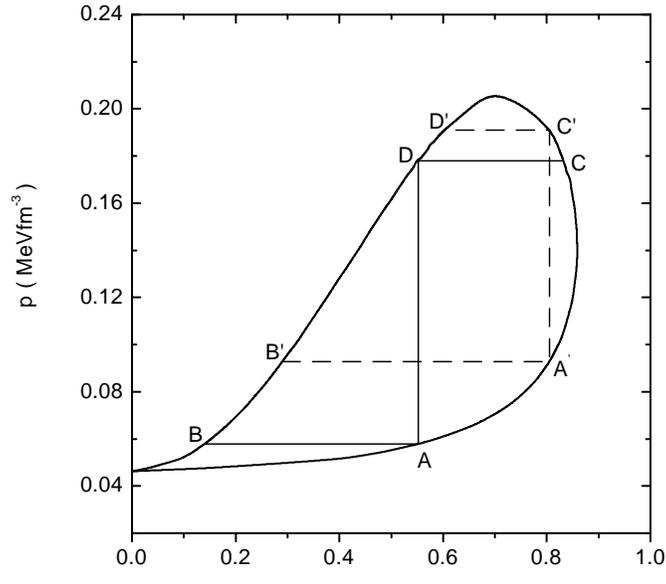,height=8cm}
}
\caption{Binodal curve at temperature $T=10$ MeV. The points A
through D and A$^\prime$ through D$^\prime$ denote two kinds of
phase transition.}
\end{figure}
\end{center}

The $\alpha$ dependence of the critical temperature is
shown in Fig.~5. The solid and dashed lines correspond to the square
root ans\"atz and the linear definition of effective nucleon mass, respectively.
$T_c$ decreases with increasing $\alpha$. For the square root
case, when $\alpha$ is less than 0.2, the decrease of $T_c$ is very
small. When $\alpha$ is larger than 0.6, $T_c$ decreases very
fast. If $\alpha$ is larger than 0.88, the system can only be in
the gas phase at any temperature. In the linear case, the critical
temperature is 2 MeV larger than that in the square root case.
The liquid-gas phase transition can occur for nuclear matter with any $\alpha$
if the temperature is lower than the critical temperature.
The critical temperature of asymmetric nuclear matter was also studied
in the Walecka model, the derivative scalar coupling model and the QMC model
\cite{Song2,Songhq} where the authors found different critical temperatures
for protons and neutrons. The lower critical temperature was chosen to be the
$T_c$ of the system as an approximation. Their results show that $T_c$ decreases
almost linearly with increasing $\alpha$, which is different from the results
shown in Fig.~5. We use the stability conditions on the free energy, conservation
laws and the Gibbs criterion for the liquid-gas phase transition of asymmetric
nuclear matter, following M\"uller and Serot \cite{Muller}. This is expected
to be better than the approximation used in the earlier discussion.

\begin{center}
\begin{figure}[hbt]
\centering{\
\epsfig{figure=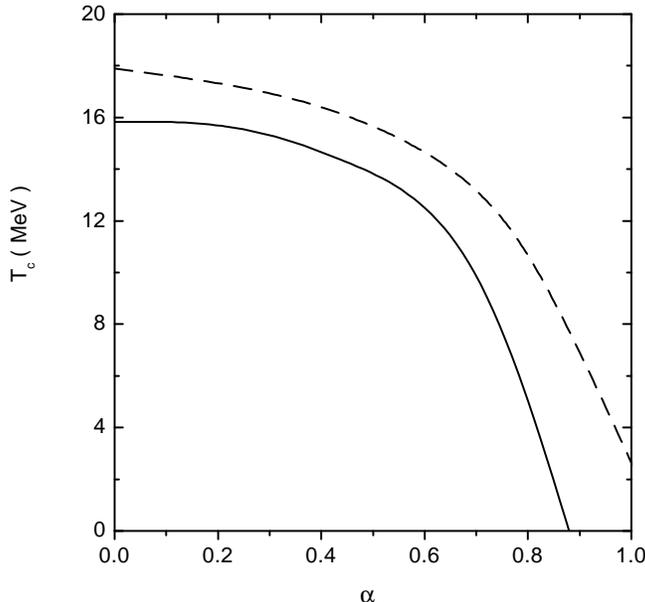,height=8cm}
}
\caption{The critical temperature $T_c$ versus asymmetric parameter
$\alpha$. The solid and dashed lines are for the square root and
linear cases, respectively.}
\end{figure}
\end{center}

\subsection{Coulomb instability}

In this subsection, we discuss the liquid-gas phase equilibrium for
finite nuclei. Compared with the case of infinite nuclear matter, the size
effect and Coulomb interaction are important for finite nuclei.
When the Coulomb interaction is considered, the chemical potential
for the protons will have an additional term
corresponding to a uniformly charged sphere:
\begin{equation}
\mu_{Coul}=\frac65\frac{Ze^2}{R},
\end{equation}
where $Z$ and $R$ are the charge number and radius of finite
nuclei. Meanwhile, the pressure also has an extra term:
\begin{equation}
p_{Coul}(\rho)=\frac{Z^2e^2}{5AR}\rho,
\end{equation}
where A is the mass number of the nucleus and $\rho$ is its density.
For a liquid droplet, the surface pressure is
\begin{equation}
p_{surf}(T,\rho)=-2\gamma(T)/R,
\end{equation}
In the above equation, $\gamma(T)$ is the surface tension suggested
by Goodman $et$ $al.$ \cite{Goodman}
\begin{equation}
\gamma(T)=(1.14\text{MeVfm}^{-2})\left[1+\frac{3T}{2T_c}\right]\left[1-\frac{T}{T_c}\right],
\end{equation}
where $T_c$ is the critical temperature for infinite symmetric
nuclear matter.
The Gibbs conditions for two phase equilibrium now becomes
\begin{equation}
p(T,\rho_L,\alpha_L)+p_{Coul}(\rho_L)+p_{surf}(T,\rho_L)=p(T,\rho_V,\alpha_V),
\end{equation}
\begin{equation}
\mu_n(T,\rho_L,\alpha_L)=\mu_n(T,\rho_V,\alpha_V),
\end{equation}
\begin{equation}
\mu_p(T,\rho_L,\alpha_L)+\mu_{Coul}(\rho_L)=\mu_p(T,\rho_V,\alpha_V).
\end{equation}
When the temperature is higher than a temperature (limiting
temperature), the above equations have no solution. The finite
nuclei cannot exist in equilibrium with the surrounding vapor.

We show in Fig.~6 the mass number dependence of the limiting
temperature, $T_{lim}$, for nuclei along the line of $\beta$-stability:
\begin{equation}
Z=0.5A-0.3\times 10^{-2}A^{5/3}.
\end{equation}
The solid and dashed lines correspond to the square root ans\"atz
and the linear definition of the effective nucleon mass,
respectively. The experimental values obtained recently by
Natowitz \cite{Natowitz1} are also plotted in the figure for
comparison. The calculated results in this model are in good
agreement with the experimental data. The limiting temperature
decreases with increasing mass number. This means that when the
temperature is higher than the limiting temperature, the heavy
nuclei will fragment to light nuclei. Two useful parameterizations
of $T_{lim}/T_c$, valid for $10\leq A\leq 208$, are $(T_{lim}/T_C)
=0.611-0.00193A+3.32\times 10^{-6}A^2$ (square root case) and
$(T_{lim}/T_C) =0.591-0.00203A+3.80\times 10^{-6}A^2$ (linear
case), which are comparable with that given in Ref.\
~\cite{Natowitz2}.
Calculations from other models also show that $T_{lim}$ decreases with
increasing mass number. Numerical results show that there is a relationship
between $T_{lim}$ and $T_c$. For larger $T_c$, the calculated $T_{lim}$ is
also larger. Therefore, the results for $T_{lim}$ found with the Walecka model
and the SK1 interaction are larger, while the results of the SKM$^*$ interaction
are smaller compared with the experiments \cite{Songhq}. The values of $T_{lim}$
calculated for both the QMC model and the effective model suggested by Furnstahl
$et$ $al.$ \cite{Zhang2,Songhq} are close to our results.

\begin{center}
\begin{figure}
\centering{\ \epsfig{figure=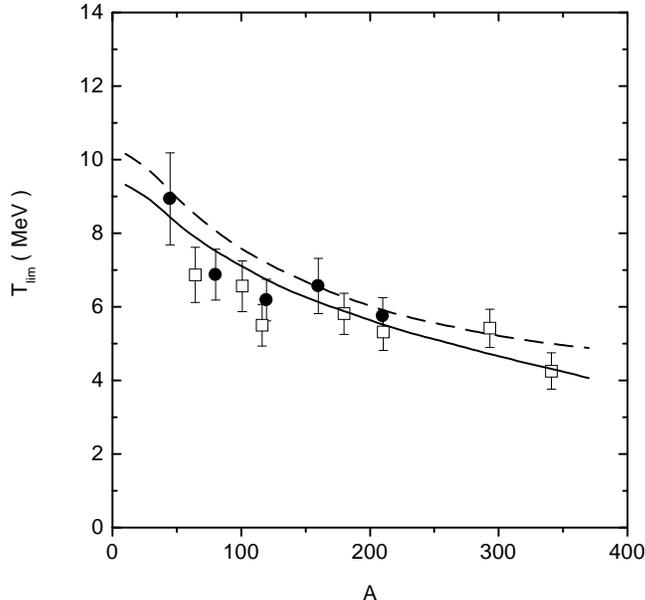,height=8cm} } \caption{The
limiting temperature $T_{lim}$ versus mass number A of finite
nuclei. The solid and dashed lines are for the square root and
linear cases, respectively. The points with error bars are from
Ref. [19]. The data derived from the double isotope yield ratio
and thermal bremsstrahlung measurements are represented by the
filled circles and open squares, respectively.}
\end{figure}
\end{center}

\section{Summary}

In this paper, we extended the chiral $SU(3)$ quark mean field model
to finite temperature and density. This model describes properties
of infinite nuclear matter, finite nuclei and hypernuclei very well
at zero temperature. The saturation properties and compression modulus
of nuclear matter are reasonable. The hyperon potentials are close to
the empirical values for hadronic matter. The results for finite nuclei
and hypernuclei are also consistent with experiment.
We therefore want to know whether this model can also describe the system
at finite temperature. The liquid-gas phase transition of
infinite nuclear matter and the Coulomb instability of finite nuclei
at finite temperature are discussed in this
model. All the parameters have been determined in earlier papers
and there is no further parameter to be adjusted.
The critical temperatures for two-phase coexistence of
symmetric nuclear matter, $T_c$, is 15.82 MeV (square root case)
and 17.9 MeV (linear case). Both of these values are close to the
recent experimental value, $16.6\pm 0.86$ MeV. $T_c$
is found to decrease with either increasing
asymmetry parameter, $\alpha$, or increasing mass number for finite nuclei.
The critical temperature, $T_c$, in
the linear case is about 2 MeV larger than that in the square root
case.
The values in both cases are in good agreement
with the experimental values found in Refs.~\cite{Natowitz1,Natowitz2}.

\bigskip
\bigskip

\section*{Acknowledgements}
This work was supported by the Australian Research Council
and by DOE contract DE-AC05-84ER40150, under which SURA operates
Jefferson Laboratory.


\begin{thebibliography}{99}
\bibitem{Suraud}  E. Suraud, C. Gr$\acute{\text{e}}$gorie and B. Tamain, Prog. Part. Nucl. Phys.
(U.K.) {\bf 23} (1989) 357.

\bibitem{Panagiotou} A. D. Panagiotou, M. W. Curtin, H. Toki,
D. K. Scott and P. J. Siemens, Phys. Rev. Lett. {\bf 52} (1984) 496.

\bibitem{Chen}  Z. Chen et al. Phys. Rev. {\bf C36} (1987) 2297.

\bibitem{Bonche} P. Bonche et al., Nucl. Phys. {\bf A436} (1985) 265.

\bibitem{Levit} S. Levit and P. Bonche, Nucl. Phys. {\bf A437} (1985) 426.

\bibitem{Besprovany} J. Besprovany and S. Levit, Phys. Lett. {\bf B217} (1989) 1.

\bibitem{Song1} H. Q. Song and R. K. Su, Phys. Rev. {\bf C44} (1991) 2505.

\bibitem{Zhang1} Y. J. Zhang, R. K. Su, H. Q. Song and F. M. Lin, Phys. Rev. {\bf C54} (1996) 1137.

\bibitem{Das} A. Das, R. Nayak and L. Satpathy, J. Phys. {\bf G18} (1992) 869.

\bibitem{Song2} H. Q. Song, Z. X. Qian and R. K. Su, Phys. Rev. {\bf C47} (1993)
2001; Phys. Rev. {\bf C49} (1994) 2924.

\bibitem{Baldo} M. Baldo and L. S. Ferreira, Phys. Rev. {\bf C59} (1999) 682.

\bibitem{Kolomietz} V. M. Kolomietz, A. I. Sanzhur, S. Shlomo and
S. A. Firin, Phys. Rev. {\bf C64} (2001) 024315.

\bibitem{Glendenning} N. K. Glendenning, Phys. Rev. {\bf D46} (1992) 1274.

\bibitem{Muller} H. M\"uller and B. D. Serot, Phys. Rev. {\bf C52} (1995) 2072.

\bibitem{Wang1} P. Wang, Phys. Rev. {\bf C61} (2000) 054904.

\bibitem{Qian} W. L. Qian, R. K. Su and P. Wang, Phys. Lett. {\bf B491} (2000) 90.

\bibitem{Pawlowski} P. Pawlowski, Phys. Rev. {\bf C65} (2002) 044615.

\bibitem{Natowitz1} J. B. Natowitz et al., Phys. Rev. {\bf C65} (2002) 034618.

\bibitem{Natowitz2} J. B. Natowitz et al., Phys. Rev. Lett. {\bf 89} (2002) 212701.

\bibitem{Zhang2} L. L. Zhang, H. Q. Song, P. Wang, and R. K. Su, Phys. Rev.
{\bf C59} (1999) 3292.

\bibitem{Carter} G. Carter, P. J. Ellis, and S. Rudaz, Nucl. Phys. {\bf A603
} (1996) 367; Erratum-ibid. {\bf A608} (1996) 514.

\bibitem{Furnstahl} R. J. Furnstahl, H. B. Tang, and B. D. Serot, Phys.
Rev. {\bf C52} (1995) 1368.

\bibitem{Papazoglou1} P. Papazoglou, S. Schramm, J. Schaffner-Bielich, H.
St\"ocker and W. Greiner, Phys. Rev. {\bf C57} (1998) 2576.

\bibitem{Papazoglou2} P. Papazoglou, D. Zschiesche, S. Schramm,
J.Schaffner-Bielich, H. St\"ocker and W. Greiner, Phys. Rev. {\bf C59}
(1999) 411.

\bibitem{Guichon} P. A. M. Guichon, Phys. Lett. {\bf B200} (1988) 235;

S. Fleck, W. Bentz, K. Shimizu and K. Yazaki, Nucl. Phys.
{\bf A510} (1990) 731;

K. Saito and A. W. Thomas, Phys. Lett. {\bf B327} (1994) 9.

P. G. Blunden and G. A. Miller, Phys. Rev. {\bf C54} (1996) 359.

H. M\"uller and B. K. Jennings, Nucl. Phys. {\bf A640} (1998) 55.

\bibitem{Kazuo} K. Tsushima, K. Saito, J. Haidenbauer and A. W. Thomas,
Nucl. Phys. {\bf A630} (1998) 691;

K. Tsushima, K. Saito and A. W. Thomas, Phys. Lett. {\bf B411} (1997) 9,
Erratum-ibid. {\bf B421} (1998) 413

\bibitem{Thomas} A. W. Thomas, S. Theberge and G. A. Miller, Phys.
Rev. {\bf D24} (1981) 216; A. W. Thomas,
Adv. Nucl. Phys. {\bf 13} (1984) 1;
G. A. Miller, A. W. Thomas and S. Theberge,
Phys. Lett. {\bf B91} (1980) 192.

\bibitem{Bentz} W. Bentz and A. W. Thomas, Nucl. Phys. {\bf A696}
(2001) 138.

W. Bentz, T. Horikawa, N. Ishii and A. W. Thomas, Nucl. Phys. {\bf
A720} (2003) 95.

H. Mineo, W. Bentz, N. Ishii, A.W. Thomas and K. Yazaki, Nucl.
Phys. {\bf A735} (2004) 482.

M. Buballa, hep-ph/0402234.

\bibitem{Toki} H. Toki, U. Meyer, A. Faessler and R. Brockmann, Phys. Rev.
{\bf C58} (1998) 3749.

\bibitem{Wang3} P. Wang, Z. Y. Zhang, Y. W. Yu, Commun. Theor. Phys.
{\bf 36} (2001) 71.

\bibitem{Wang4} P. Wang, Z. Y. Zhang, Y. W. Yu, R. K. Su and H. Q. Song,
Nucl. Phys. {\bf A688} (2001) 791.

\bibitem{Wang5} P. Wang, V. E. Lyubovitskij, Th. Gutsche and Amand
Faessler, Phys. Rev. {\bf C67} (2003) 015210.

\bibitem{Wang6} P. Wang, H. Guo, Z. Y. Zhang, Y. W. Yu, R. K. Su and H. Q.
Song, Nucl. Phys. {\bf A705} (2002) 455.

\bibitem{Guichon2} P. A. M. Guichon, K. Saito, E. Rodionov and A. W. Thomas,
Nucl. Phys. {\bf A601} (1996) 349.

\bibitem{Hackett} E. J. Hackett-Jones, D. B. Leinweber and A. W. Thomas,
Phys. Lett. {\bf B494} (2000) 89.

\bibitem{Wangcssm} P. Wang, D. B. Leinweber, A. W. Thomas and A.
G. Williams, Nucl. Phys. {\bf A744} (2004) 273.

\bibitem{Songhq} H. Q. Song and R. K. Su, J. Phys. {\bf G22} (1996) 1025.

\bibitem{Goodman} A. L. Goodman, J. I. Kapusta and A. Z. Mekjian,
Phys. Rev. {\bf C30} (1984) 851.

\end{thebibliography}
\end{document}